\documentclass[
 reprint,
 amsmath,amssymb,
 aps,
]{revtex4-2}

\usepackage{amsmath}
\usepackage{amssymb} 
\usepackage{graphicx}
\usepackage{dcolumn}
\usepackage{bm}
\usepackage{color}
\usepackage{multirow}
\usepackage{dcolumn}
\usepackage{xcolor}

\newcolumntype{L}{D{.}{.}{2,5}}

\newcommand{\G}{\mathrm{Gauss}}

\newcommand{\Esat}{(E_{\mathrm{mag}}/E_{\mathrm{kin}})_{\mathrm{sat}}}
\newcommand{\ratio}{E_{\mathrm{mag}}/E_{\mathrm{kin}}}
\newcommand{\Emag}{E_{\mathrm{m}}/E_{\mathrm{m0}}}
\newcommand{\ted}{t_{\mathrm{ed}}}
\newcommand{\solfrac}{E_{\mathrm{sol}}/E_{\mathrm{tot}}}
\newcommand{\mnras}{Mon.\ Not.\ R.\ Astron.\ Soc.\ }
\newcommand{\aap}{Astron.\ Astrophys.\ }
\newcommand{\jcap}{J.\ Cosmol.\ Astropart.\ Phys.\ }
\newcommand{\physrep}{Phys.\ Rep.\ }

\newcommand{\apjs}{Astrophys.\ J.\ S.\ }
\newcommand{\apjl}{Astrophys.\ J.\ L.\ }

\begin{document}

\title{Efficient highly-subsonic turbulent dynamo and \\ growth of primordial magnetic fields}

\author{Radhika Achikanath~Chirakkara$^{1,2,3}$}
\email{radhika.ac@students.iiserpune.ac.in}
\author{Christoph Federrath$^{2}$}
\email{christoph.federrath@anu.edu.au}
\author{Pranjal Trivedi$^{3,4,5}$}
\email{pranjal.trivedi@hs.uni-hamburg.de}
\author{Robi Banerjee$^{3}$}
\email{banerjee@hs.uni-hamburg.de}

\affiliation{$^{1}$Department of Physics, Indian Institute of Science Education and Research Pune, Dr. Homi Bhabha Road, Pune 411008, India \\
$^{2}$ Research School of Astronomy and Astrophysics, Australian National University, Canberra, ACT 2611, Australia\\
$^{3}$Hamburger Sternwarte, Universit\"at Hamburg, Gojenbergsweg 112, 21029 Hamburg, Germany.\\
$^{4}$Universit\"at Hamburg, {II}. Institut f\"ur Theoretische Physik, Luruper Chaussee 149, 22761 Hamburg, Germany.\\
$^{5}$Department of Physics, Sri Venkateswara College, University of Delhi 110020 India}

\begin{abstract}
 We present the first study on the amplification of magnetic fields by the turbulent dynamo in the highly subsonic regime, with Mach numbers ranging from $10^{-3}$ to $0.4$. We find that for the lower Mach numbers the saturation efficiency of the dynamo, $\Esat$, increases as the Mach number decreases. Even in the case when injection of energy is purely through longitudinal forcing modes, $\Esat$ $\gtrsim 10^{-2}$ at a Mach number of $10^{-3}$. We apply our results to magnetic field amplification in the early Universe and predict that a turbulent dynamo can amplify primordial magnetic fields to $\gtrsim$ $10^{-16}$ Gauss on scales up to 0.1 pc and $\gtrsim$ $10^{-13}$ Gauss on scales up to 100 pc. This produces fields compatible with lower limits of the intergalactic magnetic field inferred from blazar $\gamma$-ray observations. 
\end{abstract}

\maketitle

\section{Introduction}
Magnetic fields are ubiquitous on all scales in the Universe, from the surface of stars to galaxies to the voids in the large-scale structure of the Universe. The turbulent small-scale dynamo (SSD) amplifies small seed magnetic fields, by converting turbulent kinetic energy into magnetic energy \citep{Kazantsev1968,B&S2005}. The turbulent dynamo has a wide range of applications as it can operate in a variety of astrophysical situations and has been studied in the supersonic and transonic regime of turbulence \citep{CFetal11,Federrathetal2014ApJ}, however, it remains  unexplored in the extremely subsonic regime. This regime is important for studies on magnetohydrodynamic turbulence and is relevant for many processes in astrophysics and cosmology, including the amplification of primordial magnetic fields (PMF).

Several studies have inferred the presence of intergalactic magnetic fields (IGMFs) through  $\gamma$-ray observations of TeV blazars and have predicted a lower limit of \mbox{$10^{-16}$--$10^{-18}\,\G$} for the IGMF on Mpc scales \citep{Neronov2010,Tavecchioetal2010MNRAS,Dolagetal2011ApJ,TaylorVovkNeronov2011,Vovketal2012ApJ,Takahashietal2013ApJ,Finketal2015,FermiLATcollaboration2018}. The inferred lower bounds have been questioned due to the possible effect of plasma instabilities in the intergalactic medium \citep{Brodericketal2012ApJ}. However, recent studies have taken into account the effect of plasma instabilities in the observations and have shown that a lower limit on the IGMF can be placed from the blazar $\gamma$-ray observations \citep{Batistaetal2019MNRAS,Yanetal2019ApJ}. 

Understanding the origin of these magnetic fields is an unsolved problem. Magnetic fields can be generated during various phases in the early Universe \citep{Subramanian2016}. \citet{Sigletal1997prd} predict the generation of magnetic fields $\sim$ $10^{-29} \,\G $ at the electroweak phase transition and field strengths of $\sim$ $10^{-20}\,\G $ at the QCD phase transition. \citet{Turner&Widrow1988prd} predict magnetic fields with strengths \mbox{$\sim 10^{-34}$--$10^{-10}\,\G$} on a scale of 1 Mpc may be produced during inflation. 
Otherwise, the unavoidable presence of vorticity in the primordial plasma leads to the generation of weak 
magnetic fields in the radiation era \citep{Harrison1970,Matarrese2005}. Studies by \citep{Mourao-Roque&Lugones2013PhRvD,Mourao-Roque&Lugones2018JCAP} investigate the properties of hydrodynamic turbulence in the primordial plasma at the QCD phase transition. Upper limits of $\sim 10^{-9}$ Gauss \citep{Subramanian2016,Planck2016,Paoletti&Finelli2019,Zuccaetal2017,Chlubaetal2015,Trivedietal2014,Trivedietal2012,Shiraishietal2011,Trivedietal2010,Seshadri&Subramanian2009} and recent stricter limits of $\sim 5 \times 10^{-11}$ Gauss \citep{JedamzikSaveliev2019} have been placed on PMF from cosmic microwave background anisotropies.

The observed magnetic fields, in many cases, are orders of magnitude greater than the initially generated fields. To explain the magnitude of the observed strong magnetic fields in the voids of the Universe, \citet{Wagstaffetal2014} showed that the SSD can amplify the magnetic field seeds present in the early Universe. Turbulence in the early Universe is unavoidably generated by gravitational acceleration due to the primordial density perturbations (PDP), which gives rise to longitudinal (irrotational) driving modes. From \citet{Wagstaffetal2014}, we expect the turbulent dynamo in the early Universe to have operated under very subsonic conditions with Mach numbers ($\mathcal{M}$) $\sim$ \mbox{$10^{-5}$--$10^{-4}$}. 

Motivated by these predictions, we study the behaviour of the SSD in the very subsonic regime with a purely compressive driving of the turbulence. Furthermore, it has been shown that the SSD operating during the collapse of gas clouds in minihalos can give rise to rather strong magnetic fields during the formation of the first stars \citep{Suretal2010,CF2011b}.
\citet{Xu&Lazarian2016} present a consolidated study on the kinematic and the non-linear growth phases of the SSD. The authors discuss the dynamo mechanism during primordial star formation and in the first galaxies and find that during early star formation, magnetic fields on the Jeans scale cannot be easily generated \citep{Xu&Lazarian2016}. Thus, \citet{Xu&Lazarian2016} show that more work is needed to understand the initial generation of magnetic fields in the early Universe, which may play an important role during early star formation. 
Recent studies \citep{KrumholzFederrath2019,Shardaetal2020} have also investigated the role of magnetic fields in the formation of the first stars.

A previous study by \citet{CFetal11} has examined the properties of the dynamo as a function of the Mach number and the nature of turbulent driving. They investigated the case when the turbulent dynamo is driven solely by longitudinal modes for Mach numbers in the range $\mathcal{M} \sim$ \mbox{$0.1$--$20$}, thus not reaching sufficiently far into the very subsonic regime relevant for the amplification of PMF.
In this paper, we determine the properties of the SSD with non-helical magnetic fields in the very subsonic regime for Mach numbers in the range \mbox{$\mathcal{M}$ = $10^{-3}$--$0.4$} and for a wide range of turbulent driving conditions. 

\section{Methods}
We solve the following compressible, three-dimensional, ideal  magnetohydrodynamical(MHD) equations with the FLASH code on a periodic computational grid \citep{Fryxelletal2000,bouchut2007multiwave,bouchut2010multiwave}

\begin{equation}
\label{eqn:1}
\frac{\partial \rho}{\partial t} + \nabla \cdot (\rho \Vec{v}) = 0 
\end{equation}
\begin{equation}
\label{eqn:2}
\frac{\partial (\rho \Vec{v})}{\partial t} + \nabla \cdot (\rho \Vec{v} \otimes \Vec{v} -  \Vec{B} \otimes \Vec{B}) + \nabla p = \nabla \cdot (2\nu \rho S) + \rho \Vec{f}   
\end{equation}
\begin{equation}
\label{eqn:3}
\frac{\partial \Vec{B}}{\partial t} = \nabla \times (\Vec{v} \times \Vec{B}) + \eta \nabla^{2} \Vec{B}, 
\end{equation}
closed by the isothermal equation of state, $p_{\mathrm{thermal}} = c_{\mathrm{s}}^{2}\rho$, with constant sound speed, $c_{\mathrm{s}}$, and satisfying $\nabla \cdot \Vec{B} = 0$. 
In the above equations, $\rho$, $\Vec{v}$ and $\Vec{B}$ are the density, velocity and the magnetic field. $\nu$ and $\eta$ are the kinematic viscosity and the magnetic resistivity. $p$ is the sum of the thermal and magnetic pressure of the system $p = p_{\mathrm{thermal}} + (1/2)|\Vec{B}|^{2}$. $S$ is the traceless rate of strain tensor,  $S_{ij} = (1/2)(\partial_{i} v_{j} + \partial_{j} v_{i} ) - (1/3)\delta_{ij}\nabla\cdot\Vec{v}$, which captures the viscous interactions and $\Vec{f}$ is the turbulent acceleration field used to drive the turbulence. 

The acceleration field $\Vec{f}$, is modelled using the Ornstein-Uhlenbeck process in Fourier space \citep{CF2010}. In our simulations, we stir the turbulence continuously on large scales, i.e., wavenumbers $k (2\pi/L)=[1\dots3]$, where $L$ is the side length of the cubic Cartesian computational domain, as in previous studies \citep{CF2010,CFetal11}.
The forcing is modelled by a projection operator in Fourier space, which is defined as $\mathcal{P}^{\zeta}_{ij}(\Vec{k}) = \zeta \mathcal{P}^{\perp}_{ij}(\Vec{k}) + (1- \zeta)\mathcal{P}^{\|}_{ij}(\Vec{k})$, where $\mathcal{P}^{\|}_{ij} = k_{i}k_{j}/k^{2}$ is the curl-free (compressive) projection  and $\mathcal{P}^{\perp}_{ij} = \delta_{ij} - k_{i}k_{j}/k^{2}$ is the divergence-free (solenoidal) projection. The parameter, $\zeta$, defines the nature of the projection and lies in the range [0,1]. $\zeta = 0$ corresponds to injection of purely compressive (or longitudinal) modes in the velocity field and $\zeta = 1$ implies injection of purely solenoidal (or rotational) modes. The purely compressive forcing models the turbulent acceleration field, $\Vec{f}$, such that $\nabla \times \Vec{f} = 0$ and the purely solenoidal forcing has $\nabla \cdot \Vec{f} = 0$ \citep{CF2010}. The amplitude of the turbulent driving controls the amount of kinetic energy injected into the plasma and therefore the Mach number, $\mathcal{M} = v/c_{s}$.

We perform a systematic study wherein we vary the Mach number and the nature of the turbulent driving to determine their effects on the properties of the SSD. We run our simulations on uniform grids with $128^{3}$ cells and set up a turbulent initial seed field with an initial plasma beta, $\beta_{\mathrm{i}} \sim $ \mbox{$10^{10} - 10^{14}$}.
In addition to the above mentioned ideal-MHD simulations, we solve the non-ideal MHD equations on $256^{3}$ grid cells to estimate the effective Reynolds number (Re) and magnetic Prandtl number (Pm) of the ideal MHD simulations(see Figure~\ref{fig:machdep}). In agreement with earlier work \citep{CFetal11}, we find that $\mathrm{Re}\sim1500$ and $\mathrm{Pm}\sim2$ are good approximations for the effective Reynolds and magnetic Prandtl number in the ideal MHD simulations with $128^3$ grid cells. While in the early Universe, we expect much higher Re and Pm \citep{Schoberetal2012b}, the saturation level of the dynamo, which is our main concern, is converged to within a factor of 2 compared to the limit of very high Re and Pm \citep{Federrathetal2014ApJ}.

The stretch-twist-fold dynamo mechanism results in an exponential amplification of the magnetic energy, $\Emag$ = exp($\Gamma t$) where $\Gamma$ is the amplification rate, ${E_{\mathrm{m0}}}$ is the initial magnetic energy and $t$ is the time, normalized to the eddy-turnover time $\ted$, which is defined as ${\ted = L/(2 \mathcal{M} c_{\mathrm{s}})}$ \citep{Kazantsev1968,B&S2005}. The saturation efficiency of the dynamo, defined as the ratio of the magnetic energy to kinetic energy at saturation ($\Esat$), is a function of the Mach number and the nature of turbulent driving \citep{CFetal11}.

\section{Results}
We assign a model name to all our simulations. In the model name ‘‘$\mathrm{M}$'' stands for the Mach number and ‘‘$\mathrm{S}$'' stands for the solenoidal fraction ($\zeta$) in the driving field. For example, the model ‘‘$\mathrm{M0.001S0.1}$'' represents the simulation with $\mathcal{M} \sim 10^{-3}$ and a solenoidal fraction of $0.1$ in the turbulent driving.
We study the properties of the SSD in the subsonic regime, \mbox{$\mathcal{M} \sim$ $10^{-3}$--$0.4$}. A dynamo driven by solenoidal forcing shows a higher amplification rate and saturation efficiency, because in this case, the driving field injects vorticity directly into the plasma, which is then able to drive the stretch-twist-fold dynamo mechanism efficiently \citep{CFetal11}. However, with compressive forcing, solenoidal modes are not injected directly by the turbulent driving and the plasma might have zero initial vorticity. 

Before we present and discuss our numerical results, we briefly address the basic equation for the evolution of vorticity.
Vorticity, defined as $\vec{\omega} = \nabla \times \Vec{v}$ , follows the evolution equation \citep{Mee&Brandenburg2006}
\begin{equation}
    \label{eqn:vorticity}
    \frac{\partial \Vec{\omega}}{\partial t} = \nabla \times (\Vec{v} \times \Vec{\omega}) + \nu\nabla^{2} \Vec{\omega} + \frac{1}{\rho^{2}}\nabla \rho \times \nabla p + 2\nu \nabla \times S \nabla \mathrm{ln} \rho.
\end{equation}
The vorticity equation has the same structure as the induction equation~(\ref{eqn:3}) and can therefore give rise to an exponential growth of vorticity similar to the amplification of magnetic fields by the SSD, if the last three terms on the right hand side of equation~(\ref{eqn:vorticity}) are subdominant compared to the first term \citep{Schoberetal2012}.
Considering we start with zero initial vorticity, the baroclinic term $(\nabla \rho \times \nabla p)/\rho^{2}$ can not generate any vorticity, as the system is isothermal with the equation of state $p = c_{s}^{2}\rho$. However, if density gradients are present, then through viscous interactions, the last term on the right-hand side of equation~(\ref{eqn:vorticity}) can generate vorticity, which can then be amplified through the first term on the right-hand side of equation~(\ref{eqn:vorticity}).

\begin{figure}
    \includegraphics[width=\linewidth]{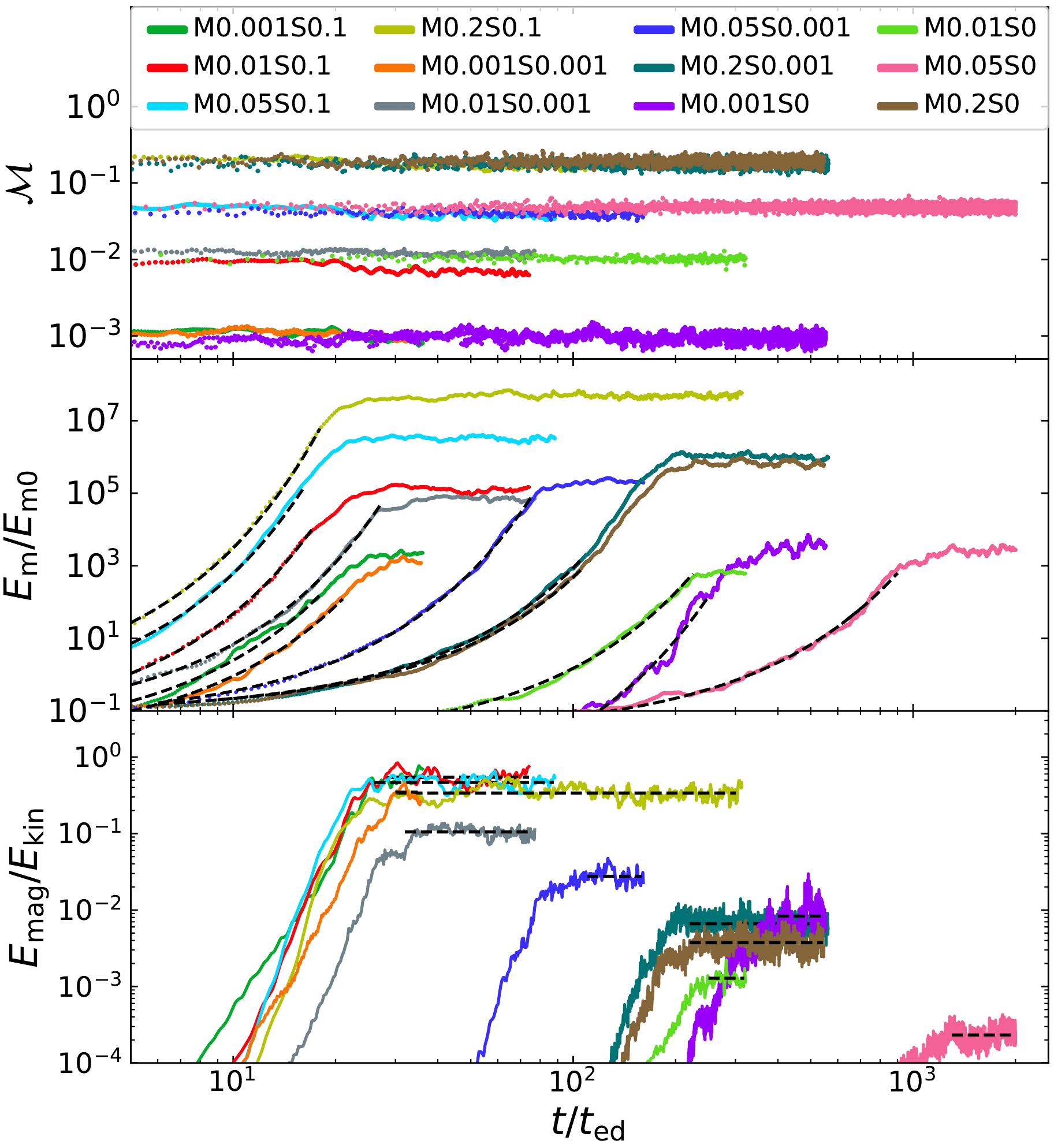}
    \caption{Mach number (top panel), magnetic energy, $\Emag$ (middle panel), and saturation level, $\ratio$ (bottom panel) as a function of time normalised to the eddy turnover time  ($\ted$) for a representative sample of our low Mach number simulation models on $128^{3}$ grid cells with solenoidal fraction of $0.1$, $0.001$ and $0$ (purely compressive) in the forcing.
    In the model name ‘‘$\mathrm{M}$'' stands for the Mach number and ‘‘$\mathrm{S}$'' stands for the solenoidal fraction ($\zeta$) in the driving field.
    The dotted lines in the middle panel show the fits for the amplification rate. The dotted black lines in the bottom panel show the fits for the saturation efficiency.}
    \label{fig:timeseries}
\end{figure}

Figure~\ref{fig:timeseries} depicts the time evolution of the Mach number, $\Emag$, and $\ratio$ as a function of time for a representative sample of our simulations (a full list of simulations is provided in the supplemental material A). 
We find that increasing the solenoidal fraction $(\zeta)$ of forcing  enhances the amplification rate of the dynamo and increases the saturation level.

\begin{figure}
    \centering
    \includegraphics[width=\linewidth]{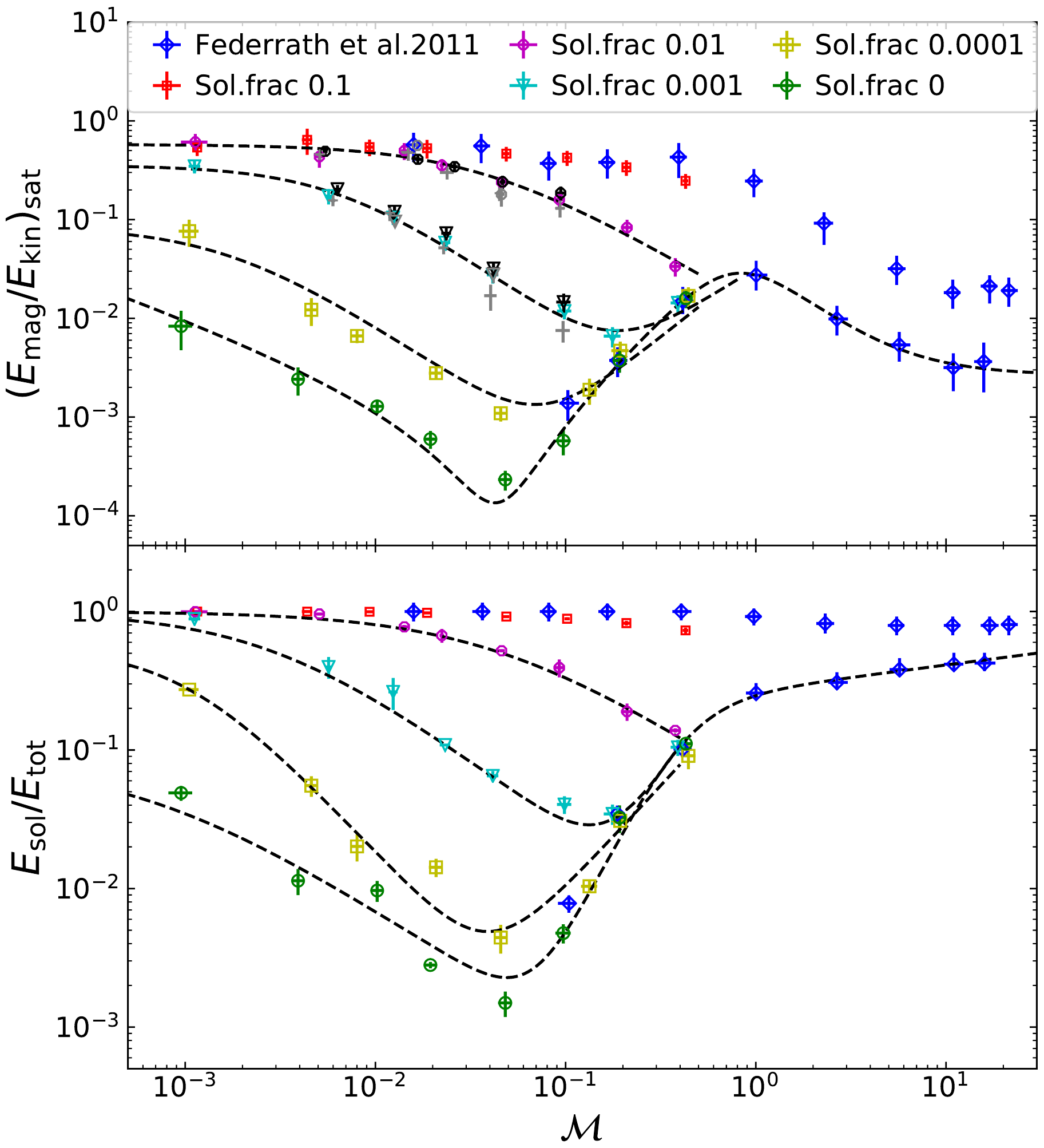}
    \caption{Saturation efficiency, $\Esat$ (top panel) and solenoidal ratio, $\solfrac$ (bottom panel) as a function of Mach number for solenoidal fraction $(\zeta)$ of 0.1, 0.01, 0.001, 0.0001 and 0 in the turbulent driving on $128^{3}$ grid cells. Dark blue (diamond) data points show purely compressive and purely solenoidal driving cases taken from Figure~3 in \citet{CFetal11}. The dotted black lines show the fits to the data to guide the eye. The black data points in the top panel correspond to the simulations done on $256^{3}$ grid cells for $\zeta = 0.01$ and $\zeta = 0.001$. The grey data points at $\mathcal{M} \sim$ 0.01 and 0.05 in the top panel correspond to simulations done on $512^{3}$ grid cells (for $\zeta = 0.01$) and $576^{3}$ grid cells (for $\zeta = 0.001$). The grey plus symbols in the top panel correspond to non-ideal MHD simulations on $256^{3}$ grid cells for $\zeta = 0.01$ and $\zeta = 0.001$ and Mach numbers in the range, $\mathcal{M} \sim$ $5 \times 10^{-3}$--$0.1$ with Reynolds number $\mathrm{Re} = 1500$ and magnetic Prandtl number $\mathrm{Pm} = 2$, i.e., these are approximately the effective Re and Pm for all the simulations on $128^{3}$ grid cells.}
    \label{fig:machdep}
\end{figure}

The saturation efficiency of the SSD and the solenoidal fraction of the kinetic energy, $\solfrac$, as a function of the Mach number and the turbulent driving, are shown in Figure~\ref{fig:machdep}. The solenoidal fraction of the kinetic energy is correlated to the amplification rate, $\Gamma$. The greater the solenoidal modes in the velocity field, the higher the vorticity of the plasma, which leads to a more efficient amplification of the magnetic energy and therefore a higher amplification rate. We find that for purely compressive driving, the amplification rate and the saturation efficiency decline with the Mach number until $\mathcal{M} \sim 0.05$. Below this Mach number, it is easier for the energy injected by the turbulence to drive rotational modes, thus generating relatively more vorticity in the plasma and increasing $\solfrac$.
The dynamo is very sensitive to the solenoidal fraction of the kinetic energy and as $\solfrac$ increases, the amplification rate and the saturation efficiency of the dynamo increase. In the very subsonic regime, both $\solfrac$ and $\Esat$ increase as the Mach number decreases.

With a solenoidal fraction of $0.1$ in the driving, we find that the saturation efficiency approaches the results from \citet{CFetal11} for purely solenoidal driving. This is also observed for the dynamo with solenoidal fractions of $0.01$ and $0.001$ in the forcing. With a solenoidal fraction of $0.0001$, we find that at $\mathcal{M} \sim 10^{-3}$, the saturation efficiency increases by an order of magnitude compared to the dynamo driven by purely compressive driving ($\zeta=0$). We also perform the low Mach number simulations with solenoidal fractions of $0.001$ and $0.01$ on $256^{3}$, $512^{3}$ and $576^{3}$ grid cells and show that the value of the saturation efficiency converges with resolution. 

The density fluctuations in the plasma decrease with the Mach number, leading to a decrease in the density gradients. This in turn enables the first term on the right hand side of equation~(\ref{eqn:vorticity}) to operate more efficiently and to generate a higher fraction of vorticity modes in the very low Mach number limit. Consequently, the kinetic energy in the rotational ($\nabla \times \vec{v}$) modes increases relative to the kinetic energy in the compressive ($\nabla \cdot \vec{v}$) velocity modes in the very subsonic regime. This causes $\solfrac$ to increase in this limit, which then leads to an efficient SSD mechanism, thereby increasing the saturation efficiency at very low Mach numbers.

\section{Applications}
Magnetic fields are unavoidably created in the primordial Universe \citep{Harrison1970} and can act as a seed for the SSD. \citet{Wagstaffetal2014} show that turbulence can be established in the early Universe between the electroweak epoch and neutrino decoupling (\mbox{$T=0.2$--$100\,\mathrm{GeV}$}), where the dissipation scale is set by neutrino damping and is $\sim 3 \times 10^{-12}$ pc in comoving coordinates at the electroweak epoch. They further describe two mechanisms for driving the turbulence in this early evolution of the Universe: 1) through velocity fluctuations generated by the PDP, and 2) through first-order phase transitions which may occur in this epoch. In the former case, the velocity fluctuations arise due to acceleration by the gravitational potential generated due to PDP and therefore are longitudinal or compressive velocity modes. They would also be driven continuously as is the case in our simulations.

Well developed turbulence together with the high magnetic Reynolds numbers and Prandtl numbers in the early Universe provides optimal conditions for the SSD to operate.  This dynamo is expected to have operated in very subsonic conditions, $\mathcal{M} \sim 10^{-4}$. In the radiation-dominated era, the relativistic equation of state, $p = \rho/3$, sets the sound speed to $c/\sqrt{3}$.

In the following discussion, we use the results obtained by \citet{Wagstaffetal2014}, where the authors estimate the magnetic fields generated by a SSD in the primordial Universe and follow the evolution of these fields to estimate the IGMF at present day. In the aforementioned work, physical quantities like the magnetic field strength and their coherence length are calculated in a co-moving frame and are evaluated at the present-day epoch. We note that the local viscosity, which determines the high Reynolds and Prandtl numbers, are set by the relativistic background in the early Universe. However, the velocity fluctuations responsible for driving the turbulence in the early Universe are non-relativistic, therefore, for our simulations of the SSD in the radiation-dominated era, the non-relativistic MHD equations are appropriate (see supplemental material B). We also note that we apply our results to the baryon-photon fluid in the early Universe prior to recombination, where using the comoving coordinates with the above mentioned relativistic equation of state is a suitable approach \citep{1998PhRvDSubramanian&Barrow,2004PhRvDBanerjee&Jedamzik,Subramanian2016,1996PhRvDBrandenburgetal,2010PhRvD_Kahniashvili_etal,2017PhRvDBrandenburgetal}.

Now, we will apply our results for the SSD in the subsonic regime to the early Universe dynamo. The turbulent dynamo action occurs on timescales substantially smaller than the expansion of the early Universe (see supplemental material C).
At $\mathcal{M} \sim 10^{-3}$ we report the saturation efficiency to be $ \sim 8.3 \times 10^{-3}$.
Taking the value of the saturation efficiency at $\mathcal{M} \sim 10^{-3}$ to be a lower bound for the early Universe dynamo, we predict the generation of magnetic fields with strengths $\gtrsim 9.1 \times 10^{-17}\,\G$ on scales up to $\lambda_{c} \sim 0.1$ pc through the dynamo action driven by PDP. 
If the SSD is driven by first-order phase transitions, we predict that the dynamo generates much higher magnetic field strengths of $\gtrsim 9.1 \times 10^{-14}\,\G$ on scales up to $\lambda_{c} \sim 100$ pc \citep{Wagstaffetal2014}. We note that these values are lower limits, as the magnetic field generated increases with the saturation efficiency, which is likely to be appreciably greater in the early Universe. We also note that these dynamo-amplified magnetic fields are well below the recent sub-nanogauss upper limits placed on PMF \citep{JedamzikSaveliev2019} but likely too weak to alleviate the Hubble tension \citep{JedamzikPogosian2020}.

The conservative estimates of the lower bounds on the IGMF from blazar $\gamma$-ray observations are \mbox{$10^{-17}$--$10^{-14}\,\G$} on scales of 0.1 pc and \mbox{$10^{-19}$--$10^{-15}\,\G$} on scales of 100 pc \citep{Finketal2015,FermiLATcollaboration2018}. The SSD mechanism driven by first-order phase transitions in the early universe can therefore explain the lower-limit on the IGMFs on scales of $\sim 100$ pc. Our important conclusion is that the dynamo mechanism driven by velocity fluctuations generated by the PDP can produce appreciable magnetic fields at shorter scales up to 0.1 pc comparable to the lower bounds on the IGMF at these scales. This raises the interesting possibility of explaining the IGMF lower bounds on these scales, without invoking beyond the standard model (BSM) physics, i.e, without requiring a first-order phase transition. In case a first-order phase transition occurs in the early Universe, it could generate stronger magnetic fields. However, the possibility of such an event in the primordial Universe is uncertain. 

These primordial fields can act as seeds for galactic dynamos and may influence the formation of the first stars \citep{Widrow2002RvMP,CF2018}. 
The Reynolds number and the Prandtl number in the early Universe are orders of magnitude higher than what we achieve in our simulations. In this limit, the growth rate increases with the Reynolds number as $\Gamma \propto Re^{1/2}$ \citep{Schoberetal2012}. Therefore, the growth rate of the early-Universe dynamo will be much higher than what is predicted from our simulations \citep{Federrathetal2014ApJ,Wagstaffetal2014}.

\section{Conclusions}
In this exploratory study of the highly subsonic MHD regime, we find that the small-scale dynamo amplifies magnetic fields efficiently for all the turbulent forcing models we have studied and the saturation efficiency increases with decreasing Mach number in the highly subsonic limit. Our results in this previously unexplored regime may be regarded as a proof-of-concept and can have wide-ranging applications for systems governed by MHD turbulence.

The results of this study can be used to estimate the magnetic field strengths produced in the early Universe by using the purely compressively-driven dynamo model. We find the small-scale dynamo action in the early Universe can generate magnetic fields with strength greater than $\sim 10^{-16}\,\G$ on scales up to 0.1 pc  when the turbulence is forced by primordial density perturbations and field strengths greater than $\sim 10^{-13}\,\G$ on scales up to 100 pc when forced by first-order phase transitions. This prediction produces fields compatible with lower limits of the intergalactic magnetic field 
inferred from blazar $\gamma$-ray observations on these scales.

\begin{acknowledgments}
We thank Amit Seta for discussions on the SSD in the early Universe.
R.~A.~would like to thank the Australian National University for the Future Research Talent award and is grateful to the University of Hamburg and the Australian National University for hosting her during the course of this project.
C.~F.~acknowledges funding provided by the Australian Research Council (Discovery Project DP170100603 and Future Fellowship FT180100495), and the Australia-Germany Joint Research Cooperation Scheme (UA-DAAD).
R.~B. and P.~T. acknowledge support by the Deutsche Forschungsgemeinschaft (DFG, German Research Foundation) under Germany’s Excellence Strategy – EXC 2121 “Quantum Universe" – 390833306. R.~B. is also thankful
for funding by the DFG through the projects BA 3706/14-1, BA 3706/15-1, BA 3706/17-1 and BA 3706/18. Computational resources used to conduct simulations presented here were provided in part by the Regionales Rechenzentrum at the University of Hamburg.
We further acknowledge high-performance computing resources provided by the Leibniz Rechenzentrum and the Gauss Centre for Supercomputing (grants~pr32lo, pr48pi and GCS Large-scale project~10391), the Australian National Computational Infrastructure (grant~ek9) in the framework of the National Computational Merit Allocation Scheme and the ANU Merit Allocation Scheme. The simulation software FLASH was in part developed by the DOE-supported Flash Center for Computational Science at the University of Chicago.
\end{acknowledgments}

\appendix

\section*{Supplemental Material A}

We assign a model name to our simulations in which ‘‘$\mathrm{M}$'' stands for the Mach number and ‘‘$\mathrm{S}$'' stands for the solenoidal fraction $(\zeta)$ in the driving field. Tabulated below, in increasing order of $\zeta$, are the values for the Mach number $(\mathcal{M})$, solenoidal fraction in the turbulent forcing, saturation efficiency of the dynamo ($\mathrm{(E_{mag}/E_{kin})_{sat}}$), amplification rate of the magnetic energy ($\Gamma$) and the solenoidal ratio in the kinetic energy $\mathrm{(E_{sol}/E_{tot}}$) for our ideal MHD simulations on $128^{3}$, $256^{3}$, $512^{3}$ and $576^{3}$ grid cells (see table \ref{table:1}) and for the non-ideal MHD simulations with a kinetic Reynolds number, $\mathrm{Re} = 1500$ and magnetic Prandtl number $\mathrm{Pm} = 2$ on $256^{3}$ gird cells (see table \ref{table:2}).

\begin{table*}
\centering
\begin{tabular}{l c c c c c} 
 \hline
 Model (Resolution $128^{3}$)& $\mathcal{M}$ & $\zeta$ & $\mathrm{(E_{mag}/E_{kin})_{sat}}$ & $\Gamma$ ($\ted^{-1}$) & $\mathrm{E_{sol}/E_{tot}}$ \\ [0.5ex] 
 \hline
$ \mathrm{M}0.001\mathrm{S}0$ & $ (9.5 \pm 1.4) \times 10^{-4} $ & 0 & $ (8.3 \pm 3.6) \times 10^{-3} $ & $ (5.5 \pm 0.9) \times 10^{-2} $ & $ (4.9 \pm 0.6) \times 10^{-2} $ \\
$ \mathrm{M}0.005\mathrm{S}0$ & $ (3.9 \pm 0.3) \times 10^{-3} $ & 0 & $ (2.4 \pm 0.8) \times 10^{-3} $ & $ (2.1 \pm 0.2) \times 10^{-2} $ & $ (1.1 \pm 0.2) \times 10^{-2} $ \\
$ \mathrm{M}0.01\mathrm{S}0$ & $ (1.0 \pm 0.1) \times 10^{-2} $ & 0 & $ (1.3 \pm 0.2) \times 10^{-3} $ & $ (4.8 \pm 0.3) \times 10^{-2} $ & $ (9.7 \pm 1.7) \times 10^{-3} $ \\
$ \mathrm{M}0.02\mathrm{S}0$ & $ (1.9 \pm 0.2) \times 10^{-2} $ & 0 & $ (6.0 \pm 1.2) \times 10^{-4} $ & $ (2.2 \pm 0.1) \times 10^{-2} $ & $ (2.8 \pm 0.1) \times 10^{-3} $ \\
$ \mathrm{M}0.05\mathrm{S}0$ & $ (4.8 \pm 0.4) \times 10^{-2} $ & 0 & $ (2.3 \pm 0.5) \times 10^{-4} $ & $ (1.1 \pm 0.1) \times 10^{-2} $ & $ (1.5 \pm 0.3) \times 10^{-3} $ \\
$ \mathrm{M}0.1\mathrm{S}0$ & $ (9.7 \pm 0.9) \times 10^{-2} $ & 0 & $ (5.7 \pm 1.7) \times 10^{-4} $ & $ (2.3 \pm 0.2) \times 10^{-2} $ & $ (4.8 \pm 0.8) \times 10^{-3} $ \\
$ \mathrm{M}0.2\mathrm{S}0$ & $ (1.9 \pm 0.2) \times 10^{-1} $ & 0 & $ (3.7 \pm 0.9) \times 10^{-3} $ & $ (8.5 \pm 0.4) \times 10^{-2} $ & $ (3.3 \pm 0.7) \times 10^{-2} $ \\
$ \mathrm{M}0.4\mathrm{S}0$ & $ (4.3 \pm 0.3) \times 10^{-1} $ & 0 & $ (1.6 \pm 0.3) \times 10^{-2} $ & $ (2.4 \pm 0.1) \times 10^{-1} $ & $ (1.1 \pm 0.1) \times 10^{-1} $ \\
$ \mathrm{M}0.001\mathrm{S}0.0001$ & $ (1.0 \pm 0.1) \times 10^{-3} $ & 0.0001 & $ (7.6 \pm 2.3) \times 10^{-2} $ & $ (2.2 \pm 0.2) \times 10^{-1} $ & $ (2.7 \pm 0.1) \times 10^{-1} $ \\
$ \mathrm{M}0.005\mathrm{S}0.0001$ & $ (4.6 \pm 0.4) \times 10^{-3} $ & 0.0001 & $ (1.2 \pm 0.4) \times 10^{-2} $ & $ (9.2 \pm 1.2) \times 10^{-2} $ & $ (5.5 \pm 0.9) \times 10^{-2} $ \\
$ \mathrm{M}0.01\mathrm{S}0.0001$ & $ (8.0 \pm 0.6) \times 10^{-3} $ & 0.0001 & $ (6.6 \pm 1.0) \times 10^{-3} $ & $ (6.0 \pm 0.3) \times 10^{-2} $ & $ (2.0 \pm 0.4) \times 10^{-2} $ \\
$ \mathrm{M}0.02\mathrm{S}0.0001$ & $ (2.1 \pm 0.2) \times 10^{-2} $ & 0.0001 & $ (2.8 \pm 0.4) \times 10^{-3} $ & $ (6.6 \pm 0.5) \times 10^{-2} $ & $ (1.4 \pm 0.2) \times 10^{-2} $ \\
$ \mathrm{M}0.05\mathrm{S}0.0001$ & $ (4.6 \pm 0.4) \times 10^{-2} $ & 0.0001 & $ (1.1 \pm 0.2) \times 10^{-3} $ & $ (3.1 \pm 0.3) \times 10^{-2} $ & $ (4.4 \pm 1.0) \times 10^{-3} $ \\
$ \mathrm{M}0.1\mathrm{S}0.0001$ & $ (1.3 \pm 0.1) \times 10^{-1} $ & 0.0001 & $ (1.9 \pm 0.6) \times 10^{-3} $ & $ (6.4 \pm 0.5) \times 10^{-2} $ & $ (1.0 \pm 0.1) \times 10^{-2} $ \\
$ \mathrm{M}0.2\mathrm{S}0.0001$ & $ (1.9 \pm 0.2) \times 10^{-1} $ & 0.0001 & $ (4.7 \pm 1.1) \times 10^{-3} $ & $ (8.0 \pm 0.4) \times 10^{-2} $ & $ (3.1 \pm 0.3) \times 10^{-2} $ \\
$ \mathrm{M}0.4\mathrm{S}0.0001$ & $ (4.4 \pm 0.4) \times 10^{-1} $ & 0.0001 & $ (1.7 \pm 0.4) \times 10^{-2} $ & $ (2.5 \pm 0.1) \times 10^{-1} $ & $ (9.1 \pm 1.8) \times 10^{-2} $ \\
$ \mathrm{M}0.001\mathrm{S}0.001$ & $ (1.1 \pm 0.1) \times 10^{-3} $ & 0.001 & $ (3.5 \pm 0.5) \times 10^{-1} $ & $ (4.4 \pm 0.3) \times 10^{-1} $ & $ (8.8 \pm 0.3) \times 10^{-1} $ \\
$ \mathrm{M}0.005\mathrm{S}0.001$ & $ (5.7 \pm 0.2) \times 10^{-3} $ & 0.001 & $ (1.7 \pm 0.3) \times 10^{-1} $ & $ (4.8 \pm 0.2) \times 10^{-1} $ & $ (4.0 \pm 0.7) \times 10^{-1} $ \\
$ \mathrm{M}0.01\mathrm{S}0.001$ & $ (1.2 \pm 0.1) \times 10^{-2} $ & 0.001 & $ (1.0 \pm 0.2) \times 10^{-1} $ & $ (5.2 \pm 0.2) \times 10^{-1} $ & $ (2.6 \pm 0.7) \times 10^{-1} $ \\
$ \mathrm{M}0.02\mathrm{S}0.001$ & $ (2.3 \pm 0.1) \times 10^{-2} $ & 0.001 & $ (5.9 \pm 1.0) \times 10^{-2} $ & $ (3.5 \pm 0.1) \times 10^{-1} $ & $ (1.1 \pm 0.0) \times 10^{-1} $ \\
$ \mathrm{M}0.05\mathrm{S}0.001$ & $ (4.1 \pm 0.3) \times 10^{-2} $ & 0.001 & $ (2.8 \pm 0.5) \times 10^{-2} $ & $ (1.9 \pm 0.1) \times 10^{-1} $ & $ (6.5 \pm 0.5) \times 10^{-2} $ \\
$ \mathrm{M}0.1\mathrm{S}0.001$ & $ (9.9 \pm 0.9) \times 10^{-2} $ & 0.001 & $ (1.2 \pm 0.2) \times 10^{-2} $ & $ (1.5 \pm 0.1) \times 10^{-1} $ & $ (4.0 \pm 0.6) \times 10^{-2} $ \\
$ \mathrm{M}0.2\mathrm{S}0.001$ & $ (1.8 \pm 0.2) \times 10^{-1} $ & 0.001 & $ (6.6 \pm 1.5) \times 10^{-3} $ & $ (9.2 \pm 0.5) \times 10^{-2} $ & $ (3.5 \pm 0.6) \times 10^{-2} $ \\
$ \mathrm{M}0.4\mathrm{S}0.001$ & $ (3.9 \pm 0.3) \times 10^{-1} $ & 0.001 & $ (1.4 \pm 0.3) \times 10^{-2} $ & $ (1.7 \pm 0.1) \times 10^{-1} $ & $ (1.0 \pm 0.1) \times 10^{-1} $ \\
$ \mathrm{M}0.001\mathrm{S}0.01$ & $ (1.1 \pm 0.2) \times 10^{-3} $ & 0.01 & $ (6.1 \pm 1.3) \times 10^{-1} $ & $ (5.8 \pm 0.5) \times 10^{-1} $ & $ (1.0 \pm 0.0) \times 10^{+0} $ \\
$ \mathrm{M}0.005\mathrm{S}0.01$ & $ (5.1 \pm 0.2) \times 10^{-3} $ & 0.01 & $ (4.3 \pm 1.0) \times 10^{-1} $ & $ (5.5 \pm 0.3) \times 10^{-1} $ & $ (9.6 \pm 0.1) \times 10^{-1} $ \\
$ \mathrm{M}0.01\mathrm{S}0.01$ & $ (1.4 \pm 0.1) \times 10^{-2} $ & 0.01 & $ (5.0 \pm 1.0) \times 10^{-1} $ & $ (1.1 \pm 0.0) \times 10^{+0} $ & $ (7.8 \pm 0.2) \times 10^{-1} $ \\
$ \mathrm{M}0.02\mathrm{S}0.01$ & $ (2.2 \pm 0.1) \times 10^{-2} $ & 0.01 & $ (3.5 \pm 0.6) \times 10^{-1} $ & $ (8.6 \pm 0.5) \times 10^{-1} $ & $ (6.7 \pm 0.8) \times 10^{-1} $ \\
$ \mathrm{M}0.05\mathrm{S}0.01$ & $ (4.6 \pm 0.2) \times 10^{-2} $ & 0.01 & $ (2.4 \pm 0.3) \times 10^{-1} $ & $ (5.9 \pm 0.4) \times 10^{-1} $ & $ (5.2 \pm 0.1) \times 10^{-1} $ \\
$ \mathrm{M}0.1\mathrm{S}0.01$ & $ (9.3 \pm 0.6) \times 10^{-2} $ & 0.01 & $ (1.6 \pm 0.3) \times 10^{-1} $ & $ (5.7 \pm 0.3) \times 10^{-1} $ & $ (3.9 \pm 0.6) \times 10^{-1} $ \\
$ \mathrm{M}0.2\mathrm{S}0.01$ & $ (2.1 \pm 0.2) \times 10^{-1} $ & 0.01 & $ (8.3 \pm 1.6) \times 10^{-2} $ & $ (5.1 \pm 0.1) \times 10^{-1} $ & $ (1.9 \pm 0.3) \times 10^{-1} $ \\
$ \mathrm{M}0.4\mathrm{S}0.01$ & $ (3.8 \pm 0.3) \times 10^{-1} $ & 0.01 & $ (3.4 \pm 0.7) \times 10^{-2} $ & $ (2.8 \pm 0.3) \times 10^{-1} $ & $ (1.4 \pm 0.1) \times 10^{-1} $ \\
$ \mathrm{M}0.001\mathrm{S}0.1$ & $ (1.2 \pm 0.1) \times 10^{-3} $ & 0.1 & $ (5.4 \pm 1.0) \times 10^{-1} $ & $ (5.2 \pm 0.6) \times 10^{-1} $ & $ (1.0 \pm 0.0) \times 10^{+0} $ \\
$ \mathrm{M}0.005\mathrm{S}0.1$ & $ (4.4 \pm 0.3) \times 10^{-3} $ & 0.1 & $ (6.4 \pm 1.9) \times 10^{-1} $ & $ (6.1 \pm 0.4) \times 10^{-1} $ & $ (1.0 \pm 0.0) \times 10^{+0} $ \\
$ \mathrm{M}0.01\mathrm{S}0.1$ & $ (9.3 \pm 0.5) \times 10^{-3} $ & 0.1 & $ (5.4 \pm 1.0) \times 10^{-1} $ & $ (7.6 \pm 0.2) \times 10^{-1} $ & $ (1.0 \pm 0.0) \times 10^{+0} $ \\
$ \mathrm{M}0.02\mathrm{S}0.1$ & $ (1.9 \pm 0.1) \times 10^{-2} $ & 0.1 & $ (5.3 \pm 1.1) \times 10^{-1} $ & $ (7.5 \pm 0.2) \times 10^{-1} $ & $ (9.8 \pm 0.1) \times 10^{-1} $ \\
$ \mathrm{M}0.05\mathrm{S}0.1$ & $ (4.9 \pm 0.2) \times 10^{-2} $ & 0.1 & $ (4.7 \pm 0.8) \times 10^{-1} $ & $ (8.9 \pm 0.6) \times 10^{-1} $ & $ (9.2 \pm 0.1) \times 10^{-1} $ \\
$ \mathrm{M}0.1\mathrm{S}0.1$ & $ (1.0 \pm 0.0) \times 10^{-1} $ & 0.1 & $ (4.2 \pm 0.7) \times 10^{-1} $ & $ (9.1 \pm 0.4) \times 10^{-1} $ & $ (8.9 \pm 0.2) \times 10^{-1} $ \\
$ \mathrm{M}0.2\mathrm{S}0.1$ & $ (2.1 \pm 0.1) \times 10^{-1} $ & 0.1 & $ (3.4 \pm 0.6) \times 10^{-1} $ & $ (9.5 \pm 0.4) \times 10^{-1} $ & $ (8.2 \pm 0.3) \times 10^{-1} $ \\
$ \mathrm{M}0.4\mathrm{S}0.1$ & $ (4.3 \pm 0.1) \times 10^{-1} $ & 0.1 & $ (2.5 \pm 0.4) \times 10^{-1} $ & $ (8.6 \pm 0.5) \times 10^{-1} $ & $ (7.3 \pm 0.4) \times 10^{-1} $ \\
\hline
\hline
Model (Resolution $256^{3}$)& $\mathcal{M}$ & $\zeta$ & $\mathrm{(E_{mag}/E_{kin})_{sat}}$ & $\Gamma$ ($\ted^{-1}$) & $\mathrm{E_{sol}/E_{tot}}$ \\ [0.5ex] 
\hline
$ \mathrm{M}0.005\mathrm{S}0.001$ & $ (6.3 \pm 0.2) \times 10^{-3} $ & 0.001 & $ (2.0 \pm 0.2) \times 10^{-1} $ & $ (8.8 \pm 0.7) \times 10^{-1} $ & $ (6.0 \pm 0.4) \times 10^{-1} $ \\
$ \mathrm{M}0.01\mathrm{S}0.001$ & $ (1.3 \pm 0.1) \times 10^{-2} $ & 0.001 & $ (1.2 \pm 0.2) \times 10^{-1} $ & $ (8.4 \pm 0.3) \times 10^{-1} $ & $ (3.8 \pm 0.2) \times 10^{-1} $ \\
$ \mathrm{M}0.02\mathrm{S}0.001$ & $ (2.4 \pm 0.1) \times 10^{-2} $ & 0.001 & $ (7.3 \pm 1.1) \times 10^{-2} $ & $ (5.7 \pm 0.3) \times 10^{-1} $ & $ (2.0 \pm 0.4) \times 10^{-1} $ \\
$ \mathrm{M}0.05\mathrm{S}0.001$ & $ (4.2 \pm 0.3) \times 10^{-2} $ & 0.001 & $ (3.2 \pm 0.5) \times 10^{-2} $ & $ (2.7 \pm 0.1) \times 10^{-1} $ & $ (8.5 \pm 1.1) \times 10^{-2} $ \\
$ \mathrm{M}0.1\mathrm{S}0.001$ & $ (9.8 \pm 0.8) \times 10^{-2} $ & 0.001 & $ (1.5 \pm 0.3) \times 10^{-2} $ & $ (2.5 \pm 0.1) \times 10^{-1} $ & $ (5.0 \pm 0.2) \times 10^{-2} $ \\
$ \mathrm{M}0.005\mathrm{S}0.01$ & $ (5.4 \pm 0.2) \times 10^{-3} $ & 0.01 & $ (4.9 \pm 0.5) \times 10^{-1} $ & $ (1.0 \pm 0.0) \times 10^{+0} $ & $ (9.9 \pm 0.0) \times 10^{-1} $ \\
$ \mathrm{M}0.01\mathrm{S}0.01$ & $ (1.7 \pm 0.1) \times 10^{-2} $ & 0.01 & $ (4.1 \pm 0.4) \times 10^{-1} $ & $ (1.6 \pm 0.1) \times 10^{+0} $ & $ (8.5 \pm 0.6) \times 10^{-1} $ \\
$ \mathrm{M}0.02\mathrm{S}0.01$ & $ (2.6 \pm 0.1) \times 10^{-2} $ & 0.01 & $ (3.4 \pm 0.4) \times 10^{-1} $ & $ (1.4 \pm 0.1) \times 10^{+0} $ & $ (6.8 \pm 0.6) \times 10^{-1} $ \\
$ \mathrm{M}0.05\mathrm{S}0.01$ & $ (4.7 \pm 0.2) \times 10^{-2} $ & 0.01 & $ (2.4 \pm 0.4) \times 10^{-1} $ & $ (9.4 \pm 0.6) \times 10^{-1} $ & $ (5.0 \pm 0.0) \times 10^{-1} $ \\
$ \mathrm{M}0.1\mathrm{S}0.01$ & $ (9.4 \pm 0.5) \times 10^{-2} $ & 0.01 & $ (1.9 \pm 0.3) \times 10^{-1} $ & $ (8.5 \pm 0.6) \times 10^{-1} $ & $ (3.3 \pm 0.3) \times 10^{-1} $ \\
\hline
Model (Resolution $512^{3}$)&  &  & &  & \\ [0.5ex] 
\hline
$ \mathrm{M}0.01\mathrm{S}0.01$ & $ (1.7 \pm 0.0) \times 10^{-2} $ & 0.01 & $ (5.7 \pm 0.4) \times 10^{-1} $ & $ (2.7 \pm 0.2) \times 10^{+0} $ & $ (8.8 \pm 0.1) \times 10^{-1} $ \\
$ \mathrm{M}0.05\mathrm{S}0.01$ & $ (4.6 \pm 0.2) \times 10^{-2} $ & 0.01 & $ (1.8 \pm 0.4) \times 10^{-1} $ & $ (1.4 \pm 0.1) \times 10^{+0} $ & $ (5.7 \pm 0.3) \times 10^{-1} $ \\
\hline
Model (Resolution $576^{3}$)&  &  & &  & \\ [0.5ex] 
\hline
$ \mathrm{M}0.01\mathrm{S}0.001$ & $ (1.3 \pm 0.1) \times 10^{-2} $ & 0.001 & $ (9.5 \pm 1.2) \times 10^{-2} $ & $ (1.3 \pm 0.2) \times 10^{+0} $ & $ (3.1 \pm 0.1) \times 10^{-1} $ \\
$ \mathrm{M}0.05\mathrm{S}0.001$ & $ (4.2 \pm 0.3) \times 10^{-2} $ & 0.001 & $ (2.8 \pm 0.5) \times 10^{-2} $ & $ (4.9 \pm 0.7) \times 10^{-1} $ & $ (8.1 \pm 1.1) \times 10^{-2} $ \\
\hline
\end{tabular}
\caption{Table of all the ideal-MHD simulations with the corresponding Mach number $(\mathcal{M})$, solenoidal fraction $(\zeta)$ in the forcing of turbulent driving, saturation efficiency of the dynamo ($\mathrm{(E_{mag}/E_{kin})_{sat}}$), amplification rate of the magnetic energy ($\Gamma$) and the solenoidal ratio in the kinetic energy $\mathrm{(E_{sol}/E_{tot}}$).}
\label{table:1}
\end{table*}

\begin{table*}
    \centering
    \begin{tabular}{l c c c c c}
\hline
 Model (Resolution $256^{3}$)& $\mathcal{M}$ & $\zeta$ & $\mathrm{(E_{mag}/E_{kin})_{sat}}$ & $\Gamma$ ($\ted^{-1}$) & $\mathrm{E_{sol}/E_{tot}}$ \\ [0.5ex] 
\hline
$ \mathrm{M}0.005\mathrm{S}0.001$ & $ (6.0 \pm 0.2) \times 10^{-3} $ & 0.001 & $ (1.6 \pm 0.2) \times 10^{-1} $ & $ (6.9 \pm 0.4) \times 10^{-1} $ & $ (5.6 \pm 0.4) \times 10^{-1} $ \\
$ \mathrm{M}0.01\mathrm{S}0.001$ & $ (1.2 \pm 0.1) \times 10^{-2} $ & 0.001 & $ (1.2 \pm 0.2) \times 10^{-1} $ & $ (5.2 \pm 0.4) \times 10^{-1} $ & $ (2.8 \pm 0.5) \times 10^{-1} $ \\
$ \mathrm{M}0.02\mathrm{S}0.001$ & $ (2.3 \pm 0.1) \times 10^{-2} $ & 0.001 & $ (5.2 \pm 0.7) \times 10^{-2} $ & $ (3.8 \pm 0.1) \times 10^{-1} $ & $ (2.2 \pm 0.3) \times 10^{-1} $ \\
$ \mathrm{M}0.05\mathrm{S}0.001$ & $ (4.0 \pm 0.3) \times 10^{-2} $ & 0.001 & $ (1.7 \pm 0.5) \times 10^{-2} $ & $ (1.1 \pm 0.0) \times 10^{-1} $ & $ (8.4 \pm 0.8) \times 10^{-2} $ \\
$ \mathrm{M}0.1\mathrm{S}0.001$ & $ (9.7 \pm 0.8) \times 10^{-2} $ & 0.001 & $ (7.5 \pm 1.9) \times 10^{-3} $ & $ (8.9 \pm 0.6) \times 10^{-2} $ & $ (5.8 \pm 1.3) \times 10^{-2} $ \\
$ \mathrm{M}0.005\mathrm{S}0.01$ & $ (5.1 \pm 0.4) \times 10^{-3} $ & 0.01 & $ (4.6 \pm 0.7) \times 10^{-1} $ & $ (7.7 \pm 0.5) \times 10^{-1} $ & $ (9.6 \pm 0.1) \times 10^{-1} $ \\
$ \mathrm{M}0.01\mathrm{S}0.01$ & $ (1.5 \pm 0.1) \times 10^{-2} $ & 0.01 & $ (4.8 \pm 0.9) \times 10^{-1} $ & $ (1.3 \pm 0.1) \times 10^{+0} $ & $ (8.4 \pm 0.4) \times 10^{-1} $ \\
$ \mathrm{M}0.02\mathrm{S}0.01$ & $ (2.4 \pm 0.2) \times 10^{-2} $ & 0.01 & $ (3.0 \pm 0.5) \times 10^{-1} $ & $ (9.6 \pm 0.6) \times 10^{-1} $ & $ (6.9 \pm 0.4) \times 10^{-1} $ \\
$ \mathrm{M}0.05\mathrm{S}0.01$ & $ (4.5 \pm 0.2) \times 10^{-2} $ & 0.01 & $ (1.8 \pm 0.3) \times 10^{-1} $ & $ (5.7 \pm 0.3) \times 10^{-1} $ & $ (5.3 \pm 0.4) \times 10^{-1} $ \\
$ \mathrm{M}0.1\mathrm{S}0.01$ & $ (9.3 \pm 0.5) \times 10^{-2} $ & 0.01 & $ (1.3 \pm 0.3) \times 10^{-1} $ & $ (5.3 \pm 0.2) \times 10^{-1} $ & $ (3.9 \pm 0.4) \times 10^{-1} $ \\
\hline
    \end{tabular}
    \caption{Table of all the non-ideal MHD simulations with Reynolds number, $\mathrm{Re} = 1500$ and magnetic Prandtl number $\mathrm{Pm} = 2$ on $256^{3}$ grid cells with the corresponding Mach number $(\mathcal{M})$, solenoidal fraction $(\zeta)$ in the forcing of turbulent driving, saturation efficiency of the dynamo ($\mathrm{(E_{mag}/E_{kin})_{sat}}$), amplification rate of the magnetic energy ($\Gamma$) and the solenoidal ratio in the kinetic energy $\mathrm{(E_{sol}/E_{tot}}$).}
    \label{table:2}
\end{table*}

\section*{Supplemental Material B}

We simulate the non-relativistic baryon fluctuations in a relativistic background plasma, which drives the small-scale dynamo in the radiation-dominated Universe. The non-relativistic MHD equations in Minkowski space-time with the relativistic equation of state, $p = \rho/3$, can be used to model these fluctuations in the primordial plasma of the early Universe  \citep{1998PhRvDSubramanian&Barrow,2004PhRvDBanerjee&Jedamzik, Subramanian2016, 1996PhRvDBrandenburgetal}. This approach has been used by studies investigating the evolution of MHD turbulence in the radiation-dominated Universe through numerical simulations \citep{2010PhRvD_Kahniashvili_etal, 2017PhRvDBrandenburgetal}.

\begin{equation}
    \frac{\partial \rho}{\partial t}+ \frac{4}{3}\nabla \cdot \left( \rho \vec{v}\right)  - \vec{E} \cdot \vec{J}=0
    \label{eqn:b1}
\end{equation}

\begin{equation}
    \frac{\partial}{\partial t}\left( \rho\vec{v}\right) +(\vec{v} \cdot \nabla)\left(\rho \vec{v}\right)+\vec{v} \, \nabla \cdot\left(\rho\vec{v}\right) =-\frac{1}{4}\nabla \rho + \frac{3}{4} \vec{J} \times \vec{B}
    \label{eqn:b2}
\end{equation}

\begin{equation}
    \frac{\partial \vec{B}}{\partial t}=\nabla \times({\vec{v}} \times \vec{B})
    \label{eqn:b3}
\end{equation}

\begin{figure}[!ht]
    \centering
    \includegraphics[width=\linewidth]{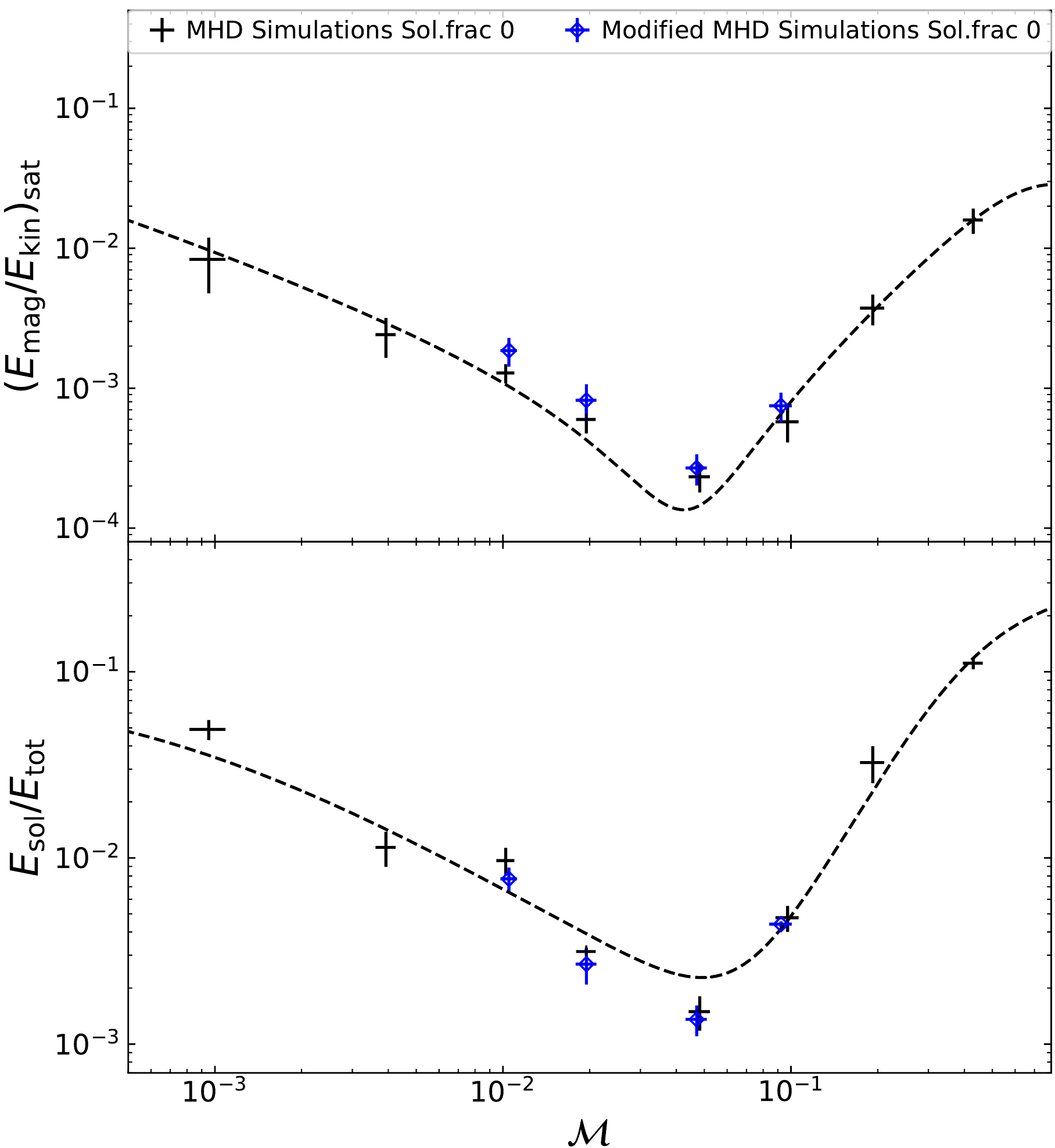}
    \caption{Saturation efficiency, $\Esat$ (top panel) and solenoidal ratio, $\solfrac$ (bottom panel) as a function of Mach number for solenoidal fraction 0 in the turbulent driving for our standard MHD simulations (on $128^{3}$ grid cells; black data points) and MHD simulations with the modified momentum equation (on $144^{3}$ grid cells; blue data points). The dotted black lines show the fits to the MHD simulation data to guide the eye.}
    \label{fig:1}
\end{figure}

The relativistic ideal MHD equations for non-relativistic velocity fluctuations in co-moving coordinates are equations (\ref{eqn:b1})-(\ref{eqn:b3}), where the relativistic equation of state, $p=\rho/3$, is used. These equations resemble the usual MHD equations (\ref{eqn:1})-(\ref{eqn:3}) albeit with some constant factors being introduced in the equations as the pressure of the relativistic plasma is significant compared to its energy density. We have modified the momentum equation we solve accordingly and find that the properties of the low-Mach number small-scale dynamo are consistent to within 1-sigma with the results obtained from solving the usual MHD equations (Figure \ref{fig:1}). 

\section*{Supplemental Material C}
The small-scale dynamo amplification of seed magnetic fields present in the primordial plasma occurs in the radiation-dominated early Universe. \citet{Wagstaffetal2014} discuss two mechanisms for generating turbulence in the early Universe: (i) Turbulence driven by primordial density perturbations and (ii) Turbulence from first-order phase transitions, and they show that the kinetic and magnetic Reynolds numbers in the early Universe are higher than the critical values required for dynamo action. The authors of the aforementioned study assume a Kolmogorov spectrum for the turbulence, however, the rapid growth and saturation of magnetic fields is attained in the early Universe independent of the nature of turbulence as the kinetic and magnetic Reynolds numbers are very high in the radiation-dominated Universe \citep{Schoberetal2012}. The rapid exponential amplification of magnetic energy is followed by a slower linear growth phase leading up to the saturation of the SSD \citep{Xu&Lazarian2016}. Neutrino damping sets the viscous dissipation scale, which is the smallest length scale at which turbulence can be maintained, in the radiation-dominated Universe (at temperatures, $T\simeq 0.2 - 100 \, \mathrm{ GeV}$). The ratio of the physical timescales for the expansion (Hubble) time, $\tau_{H}$, to the eddy turn-over time at the neutrino damping scale, $\tau_{l}$, for the primordial density perturbations are

\begin{gather*}
\frac{\tau_{H}}{\tau_{l}} = \frac{1}{H} \frac{v^{\rm rms}_l}{(a\,l_{c})} \sim 2 \times 10^{4} \, (l_{c} \simeq 3 \times 10^{-12} \,\mathrm{pc,}\, T\simeq 100\, \mathrm{GeV}) \\
\mathrm{and} \,\,  \frac{\tau_{H}}{\tau_{l}}
\sim 5 \times 10^{3}(l_{c} \simeq 10^{-10} \,\mathrm{ pc, }\, \, T\simeq 15\, \mathrm{ GeV).}   
\end{gather*}

Here the physical length scale for turbulent driving $l = al_{c}$, where $l_{c}$ is the comoving length scale and $a$ is the cosmological scale factor. In the case of energy injection into the primordial plasma due to a first-order PT

\begin{gather*}
 \frac{\tau_{H}}{\tau_{l}}
\sim 3 \times 10^{7} \,\, \mathrm{for}\,\, \, T\simeq 100 \mathrm{ GeV} \,\, \mathrm{and} \\
\frac{\tau_{H}}{\tau_{l}}
\sim 10^{7} \,\, \mathrm{for}\,\, T\simeq 15 \mathrm{ GeV.}  
\end{gather*}

We note that for length scales in between the neutrino damping scale and the largest possible driving scale (given by $v_{\rm rms}\tau_{H}$), ${\tau_{H}}/{\tau_{eddy}} \propto a$ in the radiation-dominated early Universe \citep{2004PhRvDBanerjee&Jedamzik,Wagstaffetal2014}. For purely compressive turbulence driving ($\zeta = 0$), we find the amplification of magnetic energy, including the exponential and the slower linear growth phase, and the saturation of the SSD occurs after approximately $500$ turn-over times at $\mathcal{M} \sim 10^{-3}$ in our simulations with Reynolds number $Re \simeq 1500$. However, the Reynolds numbers in the early Universe are much higher than what we obtain in our simulations; $Re \simeq 10^7$ at $ T\simeq100$ GeV and $Re \simeq 10^5$ at $T\simeq15$ GeV \citep{Wagstaffetal2014}. \citet{Schoberetal2012} find the amplification rate, $\Gamma$, of SSD-amplified magnetic fields for turbulence with different velocity scaling, following $v(\ell) \sim \ell^\theta$, to be
\begin{equation}
    \Gamma \propto Re ^{(1-\vartheta)/(1+\vartheta)},
\end{equation}

in the high Prandtl number limit $Pm \gg 1$. 
Using this for Kolmogorov ($\vartheta = 1/3$) and Burgers ($\vartheta = 1/2$) turbulence, we find $ \Gamma \propto Re ^{1/2}$ and $\Gamma \propto Re^{1/3}$, respectively. 

Therefore, for a range of length scales  $(1-100)\,l$, at epochs close to $T\simeq 100\, \mathrm{ GeV}$, the exponential amplification and saturation is reached appreciably faster in the early Universe compared to what we find in our simulations at $Re \simeq 1500$ and the growth and saturation of magnetic fields from the SSD action occurs on time scales substantially smaller than the expansion of the Universe. 
\citet{Wagstaffetal2014} also derive the net-amplification factor of the primordial magnetic fields and conclude that amplification of magnetic fields until saturation is achieved by the turbulent dynamo in the early Universe. 
While the growth rates are strongly dependent on the kinetic Reynolds number, the saturation levels are not, in the large magnetic Prandtl number limit, $Pm > 1$ \citep{Federrathetal2014ApJ}. Thus, the saturation levels we find in our simulations provide reasonable estimates of the saturation levels that would be obtained in the early Universe, where $Pm \gg 1$.

\end{document}